# Theoretical Model on Meridian Conduction

Liaofu　Luo　　　　Fengqi Bao

（Inner Mongolia University，　Hohhot，　China　010021）

**Abstract**

The physical mechanism on meridians (acupuncture lines) is studied and a theoretical model is proposed. The meridians are explained as an alternating system responsible for the integration and the regulation of life in addition to the neuro-humoral regulation. We proposed that the meridian conduction is a kind of mechanical waves (soliton) of low frequency along the slits of muscles. The anatomical- physiological and experimental evidences are reviewed. It is demonstrated that the stabilization of the soliton is guaranteed by the coupling between muscle vibration and cell activation. Therefore the propagation of mechanical wave dominates the excitation of cell groups along the meridian.  The meridian wave equations and its solution are deduced and how these results can be used in studying human healthy is briefly discussed .

## 1　Physics of meridian conduction

There have been a lot of attempts to explain the various manifestations of acupuncture lines (meridians) [1]-[4].　People have tried to interpret the meridian manifestations from the point of neuro-humoral regulation since the neuro-humoral system was regarded as the only regulation system in the human body based on modern physiology and medicine. However, from modern anatomy there has not been discovered any morphological evidence of special intra-corporal channel corresponding to the acupuncture lines in the traditional Chinese medicine (TCM). The fourteen meridian lines differ largely from the topological distributions of nerves and especially, even in the case of nervous conduction block and denervation, the acupuncture effects occur as well as usual. Moreover, with the effectiveness of the acupuncture treatment, the velocity of sensory transmission along the meridian in the perception of meridian-sensitive persons is reported much lower than that of nervous conduction. It seems that, in addition to the neuro- humoral regulation, there exists an alternating system responsible for the integration and the regulation of the life, giving important pathways of message to realize the communication and functional connection and the balance of different parts of human body.

According to TCM each meridian line is divided into two parts – the pathway with acupuncture points (a.p.) and without a.p. . The pathways with a.p. are distributed mainly on limbs and head and face, while the other pathways without a.p. are collateral, located near viscera. We proposed [5]-[8] that the pathways with a.p are networks composed of a special kind of slits of tissues (muscles principally) and the acupuncture points are the sensitive nodes in the networks. The meridian conduction is essentially a kind of mechanical wave of low frequency. That is, as a needle inserted in some acupuncture point the sensation of meridian conduction (for example, the transmission of feeling of warmth and anesthesia, etc.) occurs along the given pathway which is related to the propagation of the alternating contraction and elongation of muscle fiber bundles in the channel. Of course, the propagation of muscle vibration is not only a mechanical process, but also a neurophysiological and biochemical process since the muscle activity is carried out under the innervation of the autonomic nerve and the energy necessary for vibration is provided by biochemical process. Therefore, the frequency amplitude and waveform of the mechanical wave contains the life information of human body. The above mechanism on the transmission of biological information is supposed universal for the meridians with a.p. However, there is no mechanical wave in the pathways of meridians without a.p.,[2][9] Thus the mechanism of information transmission in the pathways without a.p. is quite different from that of meridians with a.p.. The pathways of meridians without a.p. connect soma with viscera and they transmit information through autonomic nervous system and body fluid system.

**The anatomical evidences on the existence of meridians**

1. There are 14 meridians (with a.p.) in human body [2], namely lung meridian (across upper limb and hand), large intestine meridian (across hand and upper limb), stomach meridian (across lower limb and foot), spleen meridian (across foot and lower limb) , heart meridian (across upper limb and hand), small intestine meridian (across hand and upper limb), bladder meridian (across lower limb and foot), kidney meridian (across foot and lower limb), pericardium meridian (across upper limb and hand), triple energizer meridian (across hand and upper limb), gall bladder meridian (across lower limb and foot), liver meridian (across foot and lower limb), governor vessel and conception vessel.  By comparing the meridian map with the atlas of human anatomy we find the pathways of meridians with a.p are distributed mainly on limbs torso and head.[2][10] A large part of meridians travels through the slits of muscles and others along the fibers of long muscle. For example, six slits of muscles on the forearms correspond exactly to six pathways in 14 meridians with a.p..

2. The distribution density of acupuncture points is high on the part with short and large muscle mass or with complex muscle direction. For example, the statistics shows that there are 62.5% acupuncture points are located in the slits of several muscles and others located in or at the beginning and end of muscles and tendons. [11].

3. Following traditional Chinese medicine 12 of 14 meridian lines have collaterals near viscera and these collaterals are pathways without a.p.. About 2/3 of the pathways without a.p are located at the lower part of neck and axillary and chest sides, others located at lumbosacral part. They are coincided with the autonomic nervous pathways that connects the central nervous system and visceral organs.[10].

4．It was reported that the acupoints are composed of collagen fiber that has a light transmittance of nearly 100% for far-infrared rays of 9-20 μm in the direction of meridian.[12]  The anisotropy of light transmittance supports the muscle-slit assumption of meridian structure.

**The physiological evidences on the meridian conduction**

1. The observed velocities of sensory transmission along the meridian are different for different bodies, but most of them are in vicinity of 1 to 10 cm/sec.[4][9]  The values agree well with the frequencies of contraction of muscle.  From neurophysiology the frequency of contraction is determined principally by delays in the synaptic and neuromuscular transmission (from about 1 ms to 10-20 ms) and the latent period necessary for the transformation of chemical energy to mechanical energy (typical value 10 ms).[13][14]  On the other hand, the diameter of a motor unit is about 0.1 to several mm and the length of a fiber 0.3 to 4 cm. Taking 0.3cm as the wave length and 30 ms as the frequency we obtain phase-velocity 10 cm/sec for the mechanical wave. Of course, the group-velocity of wave-packet may be smaller than the phase-velocity. Moreover, sometimes the transmission is observed in two directions along the meridians, reflected from some acupuncture points as the propagation is blocked by pressure. These facts are also in agreement with the assumption of mechanical wave on the meridian conduction.

2. The shape of the acupuncture line is belt-like. The width of belt is determined by the diameter of a motor unit of the muscle fiber bundle. The observed width of the belt is in the order of millimeter, in agreement with above estimate.[4]

3. The transmission velocity is sensitive to temperature. The time needed for muscle contraction is the synaptic delays, the neuromuscular transmission and the latent period of chemical energy transformation. Both these factors are temperature-dependent. For example, the neuromuscular transmission velocity increases by a factor 2.14 as temperature increases $10^{o}$C.[13][14] This thermo-coefficient can be used to explain the observed temperature dependence of meridian transmission [4].

4. A pressure of the order of $Kg/cm^2$ imposed on the skin where the meridian line runs across would cause the conduction block. The

pressure is in the same order of the stress occurring in the isomeric contraction of muscles. However, if three times pressure is imposed on 2 cm away from the sensor line the conduction block will not appear on the far side of the pressure point.[4] All these facts can be explained by the proposed mechanism on meridian conduction.

5. After the needle insertion, accompanying the sensory propagation along the meridian the phenomenon of sweating and feeling numb and sour was reported for the meridian-sensitive persons.[4] It can be explained by the connection of meridian with the autonomic nerves. The phenomenon is essentially the autonomic-nerve-sensitivity which easily occurs for persons with lower threshold of stimulus.

6. There exists an upper bound of frequency of stimulus which causes the sensory transmission along the acupuncture lines [4]. The upper bound is determined by the synaptic and neuromuscular transmission delays (about 10-20 ms) and by the latent period of the energy transformation (10 ms), about 30ms in total. The estimate is in agreement with experimental upper bound of 30 Hz. In connection with this, the refractory period is observed in which any stimulus can only produce local sensation but not the transmission along a line.[4] The phenomenon can be explained by the fatigue of muscle contraction – the contraction fatigue due to accumulation of metabolite lactic acid, the transmission fatigue at the nerve-muscle connection or the central fatigue in the nerve reflex arc.

The above discussions on the physics of meridians (apart from the last point 4 of anatomical evidences) were originally published in 1981.[5-7] Nowadays new experimental evidences on the *in vivo* visualization of the meridian was proposed.[15]

**Further evidences on the meridian conduction from experiments directly**

1. Recently the *in vivo* visualization of the pericardium meridian with fluorescent dyes was reported [15]. Through injection of two fluorescent dyes into the dermal layer at acupuncture points of pericardium meridian it was observed the trajectory of dye diffusion along a path of muscle slits accurately matching the meridian. While the injection at the control point did not yield any linear projections.
2. Comparison among injections at different points shows most trials of fluorescein injected at acupoint PC6 were associated with slow diffusion of dye along the path of pericardium meridian and converged on acupoint PC3.[15] It is found from the anatomy diagram that PC6 and PC3 are located at two ends of the interstitial space of muscle fibers [10]. The vibration of muscle fibers on both sides of the slit may squeeze the dye and make it flow in the gap. Furthermore, the lines of fluorescence points began to form 10 to 60 minutes after injection for fluorescein sodium and 6 minutes to 1 hour for ICG.[15]   The onset time 10 or 6 minutes can be explained by the time needed for the dyes permeating through the muscle medium in the present mechanism.

## 2   A mathematical model on meridian conduction

We have shown that the meridian conduction is a kind of mechanical wave. How the mechanical wave is correlated to the physiological activity of human body? Why the wave can propagate a long distance in the channel? What mechanism causes the stabilization of the wave? These problems were discussed preliminary by use of the method of mathematical model.[6]–[8]   As a physical system the living things should be described dynamically in spite of its extreme complexity. Since the order parameter describes the main feature of a self-organized system, slaving the fluctuating variables and controlling the activities of each subsystem and the system as a whole [16], we shall suggest a simplified model to deduce the order-parameter equation for the system.

   Suppose the cells aligned in one-dimensional chains take two states, the resting state denoted by $|0\rangle$ and the active state at the n-th site denoted by $|n\rangle$. Introduce excitation operator $B_n^+$ and annihilation operator $B_n$. (note that although these operators are frequently used in

quantum theory, the present model is classical). The collective state is

$$|\psi\rangle = \sum_n a_n(t) B_n^+ |0\rangle \tag{1}$$

where $a_n(t)$ is the coordinate describing collective motion of the n-th subsystem. Suppose the internal excitation of cells is coupled with some variables of vibration (for example, the contraction of muscle fiber, the vibration of membrane, etc.) denoted as $Q_n$ and its conjugate momentum as $P_n$. The energy (Hamiltonian) of the system can be expressed as

$$F = \langle \psi | H | \psi \rangle \tag{2}$$

$$H = \sum_n \varepsilon_n B_n^+ B_n - J(B_{n+1}^+ B_n + B_n^+ B_{n+1}) - g\sum_n Q_n B_n^+ B_n + \sum_n \frac{1}{2}(P_n^2 + \omega_0^2 Q_n^2) \tag{3}$$

Inserting (1)(3) into (2) we obtain

$$F = \sum_n \varepsilon_n a_n^* a_n - J\sum a_n^*(a_{n-1} + a_{n+1}) - g\sum_n Q_n a_n^* a_n + h$$

$$h = \sum_n \frac{1}{2}(P_n^2 + \omega_0^2 Q_n^2) \tag{4}$$

The variables $Q_n$ and $P_n$ satisfy canonical equations

$$\frac{dQ_n}{dt} = \frac{\partial F}{\partial P_n} = P_n$$

$$\frac{dP_n}{dt} = \frac{\partial F}{\partial Q_n} = -\omega_0^2 Q_n + g a_n^* a_n \tag{5}$$

It results

$$\frac{d^2 Q_n}{dt^2} = -\omega_0^2 Q_n + g a_n^* a_n \tag{6}$$

which describes a compelled vibration under the force $g a_n^* a_n$. Because the vibrational frequency of $Q_n$ is low, assuming it much lower than $\omega_0$, we have

$$Q_n \cong g a_n^* a_n / \omega_0^2 \tag{7}$$

Inserting (7) into (4) one obtains

$$F = \sum_n \varepsilon_n a_n^* a_n - J \sum a_n^*(a_{n-1} + a_{n+1}) - \frac{g^2}{\omega_0^2} \sum_n a_n^* a_n a_n^* a_n + h \quad (8)$$

This is the very Ginzburg-Landau's free energy used to discuss the superconductivity in equilibrium.

To establish a dynamical theory it is necessary to know the momentum conjugate to the coordinate $a_n$ and $a_n^*$. By use of the relation between Lagrangian and Hamiltonian in classical mechanics one assumes the Lagrangian of the system as

$$L = \mu \sum_n a_n^* \frac{da_n}{dt} + \mu^* \sum_n a_n \frac{da_n^*}{dt} - F(a_n, a_n^*, Q_n, P_n) \quad (9)$$

The canonical momentum conjugate to $a_n$ is

$$\pi_n = \frac{\partial L}{\partial \frac{da_n}{dt}} = \mu a_n^* \quad (10)$$

Consider a system that possibly includes dissipative force. From classical mechanics the dynamical equations for the system should be given by the time evolution of $a_n$ and $\pi_n$

$$\frac{da_n}{dt} = \frac{\partial F}{\partial \pi_n} = \frac{1}{\mu} \frac{\partial F}{\partial a_n^*} \quad (11)$$

$$\frac{d\pi_n}{dt} = -\frac{\partial F}{\partial a_n} + f_n \quad (12)$$

Here $f_n$ is the dissipative force. The consistency of two equations (11) and (12) requires

$$f_n = (1+\frac{\mu}{\mu^*})\frac{\partial F}{\partial a_n} \tag{13}$$

For a conservative system, $f_n=0$, which means $\mu$ a pure imaginary

$$\mu = iv \tag{14}$$

Substituting (8) into (11) we obtain

$$\mu\frac{da_n}{dt} = \varepsilon_n a_n - J(a_{n-1}+a_{n+1}) - \frac{g^2}{\omega_0^2}|a_n|^2 a_n \tag{15}$$

In continuous approximation it reads

$$\mu\frac{\partial a(x,t)}{\partial t} = (\varepsilon - 2J)a(x,t)) - \frac{g^2}{\omega_0^2}|a(x,t)|^2 a(x,t) - Jl^2\frac{\partial^2 a(x,t)}{\partial x^2} \tag{16}$$

Here $\varepsilon = \varepsilon_n$ has been assumed and $l$ denotes the distance between two neighboring subsystems.

Eq (16) is the resulting equation which we suggest to describe the meridian conduction. In the approximation of neglecting dissipation $\mu$ is a pure imaginary; otherwise, $\mu$ is a complex. In most real cases the influence of dissipative force can be overcome through acupuncture at some points at meridian to input the excitation energy. In this case, $\mu = iv$, the equation (16) is called nonlinear Schrodinger equation and its solution $a(x,t)$ is a solitary wave with velocity $v_s$ energy E and effective mass m,

$$E = E_0 + \frac{mv_s^2}{2}$$
$$E_0 = \varepsilon - 2J - \frac{g^4}{16Jl^2\omega_0^2} \tag{17}$$
$$m = \frac{v^2}{2Jl^2}$$

The solution shows an envelope moving with velocity $v_s$ and extending a width $\Delta x \approx \dfrac{4Jl^2\omega_0^2}{g^2}$. The amplitude of $a(x,t)$ is proportional to $\sqrt{\dfrac{g^2}{2Jl^2\omega_0^2}}$.

Define $\dfrac{g^4}{16Jl^2\omega_0^2} = G$ which is an important parameter representing the coupling between vibrational variable $Q_n$ and cell state $a_n$. From (17) we found that the internal energy $E_0$ of soliton has been lowered by a quantity $G$ due to the nonlinear interaction $g^2$ term occurring in Eq (8). The coupling between variables $Q_n$ of muscle vibration and variables $a_n$ of subsystem's collective motion guarantees the stabilization of the meridian wave packet. Eq (17) shows that the excitation energy $E$ includes two parts, the internal energy $E_0$ and the velocity-related kinetic energy. The latter depends on the effective mass $m$ of the soliton which is in turn dependent of the coupling between neighboring subsystem. The greater the excitation energy $E$ is, the higher the frequency of the meridian vibration and the higher the velocity of the soliton will be. On the other hand, for given energy $E$ the narrower the width of wave pocket, the faster the speed and the more stable the wave is.

Above studies indicate that there exist two important meridian parameters correlated with the human healthy. One is the energy $E$ or kinetic energy $\dfrac{mv_s^2}{2}$ of the soliton and another is the stability and strength of the solitary wave characterized by the quantity $G$. Therefore, the simplified model opens a way of studying the human healthy from meridian conduction.

In case of μ not a pure imaginary, one can search for the wave solution of Eq (16) in the form

$$a(x,t) = q(x - v_s t)e^{i\vartheta(t)} \qquad (18)$$

As μ a negative real one found q moves as a damping oscillator in a potential with bi-minima and approaches some value $q_0$. As μ a positive real, q(t) is the time-reversal of the above solution. Both they are solution of kink-type. As μ a complex the solution q(t) is an elliptic function .[8]

## 3  Discussions

**3.1**. Why the meridian wave has important physiological meaning? Following synergetics [16], the macroscopic behavior of a self-organized system near the critical point can be described by several collective modes (order parameters) that slave the subsystems of the system. The biological organism always has several stratums of structures. For example, the lowest stratum is the biological macromolecules while the cells form the higher stratum. As a rule, the large-scale quantity changes more slowly than that of small scale. So the variables describing the high stratum are slow-varying ones. The meridian wave as rooted from the contraction of muscle fibers is related to the action of cell-groups and propagates in a lower frequency. The physiological and pathological changes of human body should have corresponding variation and modulation of meridian waves. Therefore, the meridian wave plays the role of the slow-varying dynamical variables of the system. It constitutes a member of the order parameter set of human body. From anatomy the acupuncture points near the arteries or veins are about 85% of the total number of a.p. Moreover, meridian lines have collaterals without a.p. which are near viscera and connected with soma through autonomic nervous system. Thus the meridian waves should couple with blood vessels and nerves and contain tremendous life information.

Suppose *A* is a set of slow variables of an inhomogeneous system, from Liouville theorem and by use of Zwanzig-Mori;s projection operator technique on can deduce [17][18]

$$\frac{\partial A(t,x)}{\partial t} = i\Omega A(t,x) - \int_0^t d\tau k(\tau) A(t-\tau,x) + (\textit{fluctuating force})$$

(19)

which shows that the order parameter *A* always oscillates with frequencies $\Omega$ (the eigenvalues of $\Omega$). That the infrasound of frequencies from 2 to 20 Hz has important effects on biosystem was reported in literatures[19]. It gives further corroboration that the low frequency vibration in meridian may play important role in human body.

**3.2．** There are huge individual differences in the experimental study of meridian phenomena. The reason is, in the one hand, due to the special mechanism of meridian conduction based on the muscle activity and in the other hand, due to the meridian perception dependent of nerve sensitivity. Both factors are different for individuals. Therefore, although the existence of meridian conduction and the mechanism on the transmission of biological information along meridians is universal for everyone but the successful experiments can be completed only for part of the crowd, i.e. for the meridian-sensitive persons. External Qi therapy of Qigong master is an extreme example worthy of attention. Some evidences on the modulation of infrared radiation was observed in the experiments of "outer flow of life power' (external Qi) [20][21]. It was reported that the frequency range of the signals is about several Hertz and the modulated signals may play therapeutic role for the human body of acceptors. In our point of view, the signals are modulated by the meridian wave and the Qigong master is a kind of meridian-sensitive persons who can induce powerful meridian waves.

**3.3**．The acupuncture points are the "relay station" for mechanical wave propagation where the energy for vibration can be provided through acupuncture and moxibustion to keep the solitary wave running in case of disease or other situations that require energy replenishment. Special material structure of acupoints was studied in literatures [12]. The possible correlation of acupuncture points with low-electrical resistance points was discussed [22][23]. Following clinical experience the acupuncture points are classified into several categories.[1][2] According to TCM, "what comes out is a well and what comes in is closed." The closed a.p. is often used to regulate the function of viscera.[2] The PC3 in the pericardium meridian is a closed a.p.. It is interesting that the recent fluorescent experiment proved it a converging point of dye migration.[15] The other closed acupuncture points are: . LU5 on the lung meridian, LI11 on the large intestine meridian, ST36 on the stomach meridian, SP9 on the spleen meridian, HT3 on the heart meridian, SI8 on the small intestine meridian, BL40 on the bladder meridian, KI10 on the kidney meridian, TE10 on the triple energizer meridian, GB34 on the gall bladder meridian, and LR8 on the liver meridian. The twelve meridians are activated in different times in a day and night. For example, the pericardium meridian is easily activated on 19:00-21:00. Meridians are a network in the human body. Luo acupoints are the nodes of the network that connect two meridian lines. There are 15 Luo acupoints in total in the meridian network. The PC6 is a Luo acupoint on the pericardium meridian.[2]  Therefore, the solitary wave was easily formed and fluorescent phenomenon was easily observed on the trajectory between acupoints PC6 and PC3 after fluorescein injection.[15]   We suggest that all these closed a.p. and Luo acupoints should be focused on in the future experiments.

**3.4.**   In the present mathematical model there are two important parameters describing the meridian wave，the coupling $G$ between muscle vibration and cell activation and the energy $E$ of the system. The existence of the stable solitary wave in human body is guaranteed by the coupling $G$ since the propagation of mechanical wave dominates the excitation of the subsystems along the way. The $G$ varies from person to person. For example, $G$ is large for meridian-sensitive persons. Moreover the values of $G$ are different in fourteen meridians for any individual. Set $G$ denoted by $G_i$, $i$=1-14 for an individual. For twelve meridians $G_i$ changes in a day and night with periodicity 24 hours. These parameters contain large enough information of human body. We expect that the personal physique, disease condition and immunity etc can be described through $G_i(t)$ {$i$=1,…,14}. On the other hand, biology is a dissipative system. The solitary wave must be maintained by energy input to guarantee a definite value of $E$. In the normal health of the human body the meridian wave coordinates the functions of various tissues and organs of the body, and the dissipative energy can be replenished automatically. However, as meridian blockage and dysfunction due to physiological or pathological reasons one can open up the meridians and collaterals through acupuncture and moxibustion to provide energy for vibration and keep the solitary wave running normally.